# Characterization of the TRIGA Mark II reactor full-power steady state

Antonio Cammi, Matteo Zanetti

*Politecnico di Milano - Department of Energy*
*CeSNEF (Enrico Fermi Center for Nuclear Studies)*
*via La Masa 34 - 20156 Milano (Italy)*

Davide Chiesa, Massimiliano Clemenza, Stefano Pozzi, Ezio Previtali, Monica Sisti

*University of Milano-Bicocca*
*Physics Department "G. Occhialini" and INFN section*
*Piazza dell'Ateneo Nuovo, 20126 Milan (Italy)*

Giovanni Magrotti, Michele Prata, Andrea Salvini

*University of Pavia*
*Applied Nuclear Energy Laboratory (L.E.N.A.)*
*Via Gaspare Aselli 41, 27100 Pavia (Italy)*

**Abstract**

In this work, the characterization of the full-power steady state of the TRIGA Mark II nuclear reactor of the University of Pavia is performed by coupling Monte Carlo (MC) simulation for neutronics with "Multiphysics" model for thermal-hydraulics.
Neutronic analyses have been performed starting from a MC model of the entire reactor system, based on the MCNP5 code, that was already validated in fresh fuel and zero-power configuration (in which thermal effects are negligible) using the available experimental data as benchmark.
In order to describe the full-power reactor configuration, the temperature distribution in the core is necessary. To evaluate it, a thermal-hydraulic model has been developed, using the power distribution results from MC simulation as input. The thermal-hydraulic model is focused on the core active region and takes into account sub-cooled boiling effects present at full reactor power. The obtained temperature distribution is then introduced in the MC model and a benchmark analysis is carried out to validate the model in fresh fuel and full-power configuration.
The good agreement between experimental data and simulation results concerning full-power reactor criticality, proves the reliability of the adopted methodology of analysis, both from neutronics and thermal-hydraulics perspective.

**1. INTRODUCTION**

The TRIGA (Training Research and Isotope production General Atomics) Mark II is a research reactor designed and manufactured by General Atomics. It is a pool-type reactor cooled and partly moderated by light water, with the fuel consisting of a uniform mixture of uranium (8%wt, enriched at 20%wt in $^{235}$U) and zirconium hydride (ZrH), which provides neutron moderation inside the fuel itself. This particular composition has a large, prompt negative thermal coefficient of reactivity, meaning that as the temperature of the core increases, the reactivity rapidly decreases.

The TRIGA Mark II reactor installed at the University of Pavia is licensed for operating at 250 kW full-power steady state conditions. The core shape is a right cylinder 44.6 cm in diameter with the fuel elements distributed along five concentric rings. Fuel elements are 72.06 cm in length and 3.76 cm in diameter. The effective dimensions of the fuel material are 35.60 cm in length and 3.58 cm in diameter. The reactivity of the system is handled by means of three control rods, named SHIM, Regulating (REG) and Transient (TRANS).
The TRIGA Mark II reactor of the University of Pavia was brought to its first criticality in 1965 and since then it has been used for several scientific and technical applications such as production of radioisotopes, nuclear activation analysis, development of boron neutron capture therapy in the medical field and reactor physics studies, thanks to the possibility of performing experimental measurements for the validation of reactor modelling codes.
In this respect, a research activity was initiated since a few years, with the aim of implementing reliable simulation models for the analysis of reactor neutronics and thermal-hydraulics with an integrated approach (Cammi et al., 2013,

Borio et al., 2014a, Chiesa et al., 2014a, Chiesa et al., 2014b).

In this framework, a complete 3-D Monte Carlo model for neutronics was developed to simulate the reactor configuration at the first startup (1965), characterized by fresh fuel and zero-power (10 W). This is the reactor simplest configuration, because the fuel is not heavily contaminated with fission reaction products and its original composition is known from the data sheets which accompanied the fuel element shipments. Moreover, at zero-power the fuel can be considered in thermal equilibrium with the water of the pool and, thus, can be simulated at room temperature.

After the good agreement obtained in the benchmark analysis of the zero-power configuration (Borio di Tigliole et al., 2010; Alloni et al. 2014), we aim to characterize the full-power steady state operating conditions of the reactor in its original configuration. The main challenge in this kind of analysis is the evaluation of the fuel-moderator temperature distribution within the core. Indeed, only two fuel elements in the reactor were equipped with thermocouples for the measurement of the fuel temperature and the available data are not sufficient to characterize the temperature profiles in the axial and radial directions of the core.

Therefore, an integrated analysis which couples neutronics and thermal-hydraulics models is required.

The strategy pursued to characterize the TRIGA reactor at full-power is developed according the following steps (which are summarized in Figure 1):
- the MC model for neutronics is used to evaluate the power distribution in the core, starting with a uniform temperature distribution at 300 K;
- the power data are used as input for the thermal-hydraulics model, which provides an evaluation of the temperature distribution;
- the temperature distribution is introduced in the neutronics model to obtain more accurate results, which account for the thermal effects on reactivity and neutron fluxes.

In principle, these steps can be recursively repeated updating each time the power and temperature distributions until the results converge to the desired accuracy.

Finally, the thermal-hydraulics model can be assessed by comparing the simulated fuel temperature with thermocouple data. Furthermore, the neutronics model can be benchmarked by computing the multiplication factor ($k_{eff}$) corresponding to the criticality configurations which were recorded in 1965, when the reactor was brought to the full-power steady state for the first time.

This paper is subdivided in 6 parts: section 2 presents the neutronics model, section 3 to 5 deal with the thermal-hydraulics modelling and results, while the last section discusses the benchmark analysis of the full-power model for neutronics.

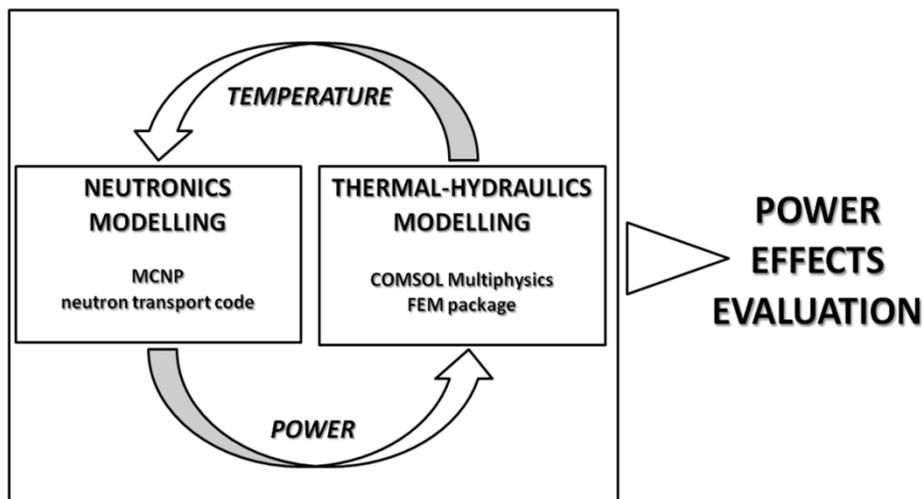

Figure 1 - Conceptual scheme of the coupling between the neutronics and thermal-hydraulics models.

## 2. NEUTRONICS MODELING

Neutronics modeling has been addressed with the Monte Carlo approach, using MCNP5 (X-5 Monte Carlo Team, 2003) for computations. The MCNP (**M**onte **C**arlo **N**-**P**article) transport code represents one of the reference codes in the field of nuclear reactor analysis. Among those, it was chosen for its capability to simulate the transport effects with a continuous-energy cross sections treatment and to model arbitrary three-dimensional configurations of materials.

In the recent years, thanks to the constant interaction with General Atomics, it was possible to obtain very detailed data concerning the most important core components of the TRIGA Mark II reactor installed at the University of Pavia. Particularly, focused searches were performed to collect precise information about the geometry and the materials of the fuel elements and control rods. In this way, it was possible to obtain a very detailed model of the reactor core (Alloni et al., 2014).

In order to simulate the neutronics of the reactor at 250 kW, the fuel-moderator temperature distribution must be described in the MC model. Since the neutron-induced interactions are affected by temperature effects, the neutronics codes, like MCNP, include thermal treatments and specific neutron cross sections to simulate the materials at different temperatures. ENDF/B-VII cross section libraries (Chadwick et al., 2011) are used.

In particular, for thermal systems such as TRIGA reactors, the S($\alpha$,$\beta$) thermal treatment (Bischoff et al., 1972) must be applied to correctly simulate the low-energy neutron interactions with the ZrH lattice and the $H_2O$ molecules. An important characteristic of neutron inelastic scattering from ZrH is that the scattered neutrons tend to exchange energy *quanta* with the lattice, whose fundamental vibrational energy is around 0.137 eV. As the temperature increases, the excited vibrational lattice levels become more populated and, as a consequence, the probability that a neutron receives a *quantum* of energy is higher. As a result, the neutron flux spectrum is less thermalized and the system reactivity decreases, implying a negative thermal coefficient.

In order to evaluate the temperature map within the fuel-moderator, the power distribution must be supplied to the thermal-hydraulic model of the reactor core. This task was performed by exploiting the MNCP model developed for the 1965 reactor configuration at zero-power. With the Monte Carlo approach, the neutron fluxes and, thus, the fission rate densities can be evaluated in the different volumes which are used to describe the geometry of the system: in this way, a discretized power distribution can be obtained through the MCNP model. Particularly, we decided to subdivide each fuel element into 8 sections (each 4.5cm high), to take into account the neutron flux variation along the vertical axis. Tallies were used to estimate the average fission rate ($R_f$) in each fuel volume, according to the following formula:

$$R_f = N_{235} \int \varphi(E) \Sigma_f(E) dE \qquad (1)$$

where $N_{235}$ is the number of $^{235}U$ atoms per fuel section, $\varphi(E)$ the neutron flux and $\Sigma_f(E)$ the microscopic fission cross section. The normalization of the neutron flux was determined starting from the total number of neutrons which are produced per unit time in the reactor when it operates at 250 kW. The power release was subsequently estimated by multiplying $R_f$ by the effective energy released per $^{235}U$ fission (192.9 MeV). The results are presented in Tab. 1, where -for brevity- the average power values of the fuel elements belonging to the same ring are reported.

This approach involves some approximations, like the discretization of the fuel volumes and the use of a neutronic simulation model with all temperatures set to 294 K. However, by interpolating the data referring to the different fuel sections, it was obtained a sufficiently good description of the power distribution in the core. On the other hand, for which concerns the approximation related to the room temperature model, in principle it would be possible to recursively repeat the analysis updating each time the temperature profile in the MCNP model. However, some preliminary tests in which the temperature was increased in the inner core rings showed that the power spatial distribution is not significantly affected by temperature variations.

| Fuel section | Released Power (W) | | | | |
|---|---|---|---|---|---|
| | Ring B | Ring C | Ring D | Ring E | Ring F |
| 1 | 561 | 485 | 427 | 362 | 298 |
| 2 | 703 | 608 | 526 | 432 | 338 |
| 3 | 826 | 712 | 614 | 496 | 376 |
| 4 | 879 | 760 | 657 | 528 | 404 |
| 5 | 830 | 720 | 626 | 515 | 394 |
| 6 | 722 | 632 | 548 | 463 | 347 |
| 7 | 589 | 514 | 443 | 386 | 280 |
| 8 | 443 | 391 | 341 | 308 | 229 |
| **Total** | 5553 | 4824 | 4183 | 3491 | 2666 |

*Table 1 - Average power release in the 8 sections of the fuel elements (1 → 8 from top to bottom) as a function of the core ring in which they are located. The Monte Carlo statistical error component was reduced so it can be considered negligible.*

## 3. THERMAL-HYDRAULIC MODELLING

In order to evaluate the reactor thermal behaviour, in particular of fuel elements, a numerical model has been developed, taking into account sub-cooled boiling effects. The model is focused on the study of the active core only, neglecting the reactor pool. This is considered sufficient to have a reasonable solution of the thermal problem for coupling with neutronics. For this purpose the Finite Element based software COMSOL Multiphysics® (Comsol, 2012b) has been employed.

The model features a 3-D description of the active region of the core, allowing a correct characterization of the non-symmetric arrangement of the reactor. The geometry is depicted in Figure 2. In order to lower the computational requirements, the clad-gap is not represented in the geometry. Instead, a thermal resistance between the fluid and the fuel surface is considered.

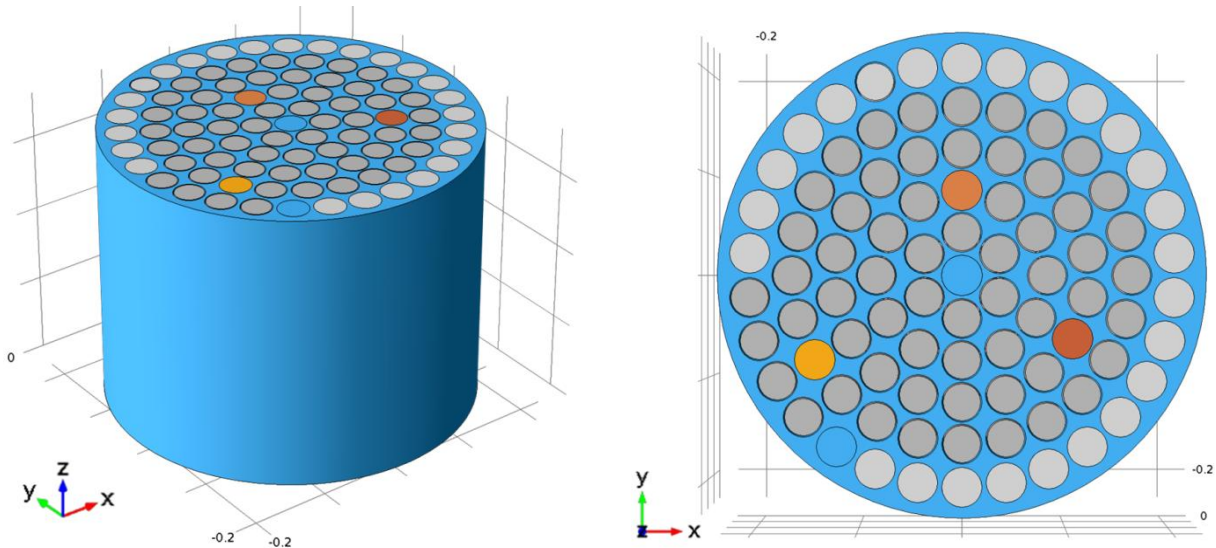

**Figure 2 - Analysed geometry, the grid spacing is equal to 0.1 m. Fuel elements are in dark grey, light grey represents graphite elements, while the control rods are represented by coloured cylinders. Each ring is identified with letters, in alphabetical order, starting from the inner channel.**

The choice of modelling the active core only entails a lack of information on the flow field behaviour. The reactor relies on natural circulation for cooling. Moreover, a secondary cooling circuit is present for radio-protection reasons, directing water towards the core top. This configuration allows confining most of the activated species in the lower part of the pool, reducing the dose on the pool top, but alters the flow behaviour in the core and the whole pool. Detailed Computational Fluid Dynamics (CFD) studies on the matter are under considerations. In the present work, the model makes use of a forced convection approach, in which the mass flow rate is imposed through the active core region from the bottom, simulating the effect of natural circulation only.

The value of the flow rate, equal to 9.3 kg/s, has been recalled from previous work (Cammi et al., 2013), to which we refer for details. Briefly, a zero dimensional natural circulation model was used to determine the equilibrium point for a given power balance. The adopted value should be considered as a rough estimate of the average effective flow-rate in the core, that is also expected to oscillate during operation. The results that will be presented in the following sections can be considered as *a posteriori* verification of the chosen value of the flow-rate. Nonetheless, the model has also been run with a reduced flow-rate (-10%), showing that the variation of the average value of the fuel temperature at full-power, that can be considered as a good indication for estimating neutronic effects, is quite low. Thus, the uncertainty on the core flow-rate at full-power can be considered negligible with respect to other uncertainties included in the model.

With the adopted flow rate value, sub-channel analysis suggests that the flow regime in the core is turbulent, with a Reynolds number ranging between 8500 and 10000. Therefore, a RANS (Reynolds Averaged Navier-Stokes) based approach has been adopted for fluid dynamics modelling. In this approach, the instantaneous quantities defined in the Navier-Stokes equations are decomposed in a time-averaged and a fluctuating part. The equations to be solved deal with the time-averaged variables, while the fluctuating quantities are found to define a stress tensor. Since the solution of a stress based problem is unpractical, the coupling between the stress tensor and the time-average velocity is represented by a quantity called turbulent viscosity, that can be determined by different models. In the present case, the "two equation" $k$-$\omega$ model available in COMSOL (Wilcox revised $k$-$\omega$ model) (COMSOL, 2012a) has been chosen for turbulence modelling. The $k$-$\omega$ model solves for the turbulent kinetic energy $k$ and for the dissipation per unit of turbulent kinetic energy $\omega$, also known as the specific dissipation rate.

Summarizing, the model is described by the following equations: mass conservation (2), momentum equation (3), *k-ω* equations (4,5) and energy equation (6):

$$\frac{\partial \rho}{\partial t} + \nabla \cdot (\rho \, \mathbf{u}) = 0 \qquad (2)$$

$$\rho \frac{\partial \mathbf{u}}{\partial t} + \rho \, (\mathbf{u} \cdot \nabla) \mathbf{u} = \nabla \cdot \left( -p\mathbf{I} + (\mu + \mu_T)((\nabla \mathbf{u}) + (\nabla \mathbf{u})^T) - \frac{2}{3} \rho k \, \mathbf{I} \right) - \mathbf{g} \rho \qquad (3)$$

$$\rho \frac{\partial k}{\partial t} + \rho \, (\mathbf{u} \cdot \nabla) k = \nabla \cdot \left( (\mu + \mu_T \sigma_k^*) \nabla k \right) + P_k - \rho \beta^* k \omega \qquad (4)$$

$$\rho \frac{\partial \omega}{\partial t} + \rho \, (\mathbf{u} \cdot \nabla) \omega = \nabla \cdot \left( (\mu + \mu_T \sigma_\omega) \nabla \omega \right) + \alpha \frac{\omega}{k} P_k - \rho \beta \omega^2 \qquad (5)$$

$$\rho \, c_p \frac{\partial T}{\partial t} + \rho \, c_p \mathbf{u} \cdot \nabla T = \nabla \cdot (\lambda \, \nabla T) + Q \qquad (6)$$

In the solid region, in which the velocity field is null, only equation (6) holds. The meaning of the symbols is reported in the Nomenclature. In the *k-ω* model, the turbulent viscosity $\mu_T$ is defined as:

$$\mu_T = \rho \frac{k}{\omega} \qquad (7)$$

The not described parameters are abridged in Appendix A.

Equations (2) to (5) are subjected to the following boundary conditions.
i) On the inlet, an uniform $U_0$ velocity corresponding to a mass flow rate of 9.3 kg/s is imposed. The turbulence variables are set to:

$$k = \frac{3}{2} (U_0 I_T)^2 \qquad (8)$$

$$\omega = \frac{k^{1/2}}{\left( \beta_0^* \right)^{1/4} L_T} \qquad (9)$$

where the turbulent intensity $I_T$ and the turbulence length scale $L_T$ are set according to the manual (Comsol, 2012a).
ii) On the outlet, a pressure of 1.5 bar is imposed while homogeneous Neumann conditions are adopted for the turbulent kinetic energy and the specific dissipation rate.
iii) The interface between the fluid and solid regions as well as the outer fluid boundary are subject to wall condition. The fluid behaviour close to a solid wall for a turbulent flow is very different if compared to the free stream case, for which turbulent models are generally developed. Therefore, analytical expressions are used to describe the flow at the walls. These expressions are known as wall functions. For the adopted wall functions, see appendix B.

The boundary conditions for equation (6) are: i) thermal insulation on the top and bottom of the solids and on the outer boundary of the fluid; ii) fixed temperature of 25 °C on the fluid bottom; on the core top, an outflow condition corresponding to thermal insulation is imposed; iii) finally, the gap-cladding is modelled as thermal resistances:

$$-\hat{\mathbf{n}}_h \cdot (-\lambda_s \nabla T_h) = -\lambda_s (T_h - T_{h^*})/d_s \qquad (10)$$

$$\lambda_s = d_s / \sum_j d_{s,j} / \lambda_{s,j} \qquad (11)$$

$$d_s = \sum_j d_{s,j} \qquad (12)$$

where the h subscript indicates a point and h* its dual on the other surface. Values of $d_s$ for the clad and the gap are 0.76 and 0.14 mm (Cammi et al., 2013). Values of $\lambda_s$ for the clad and the gap are 228.00 and 0.05 W·m$^{-1}$·K$^{-1}$, respectively. Such modelling does not take into account axial conduction in the cladding, however, such effect is negligible when compared to forced convection heat transfer at full flow rate.

The boundary conditions are summarized in Figure 3.

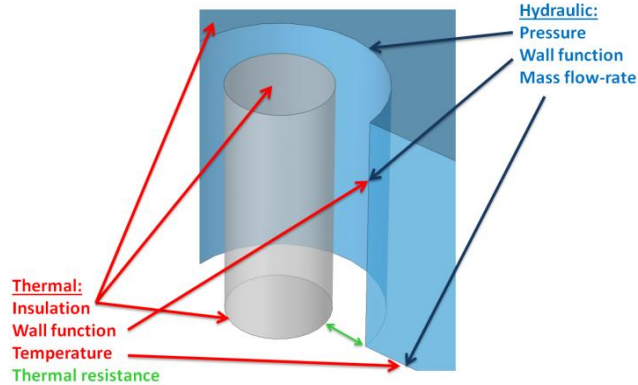

**Figure 3 - Conceptual scheme for boundary conditions. Not in scale.**

The power source in equation (6) was determined by the neutronic model described in section 2. In order to provide to the code a continuous function, the data reported in Table 1 have been interpolated with fifth order polynomials. The water properties are taken from the COMSOL material library (COMSOL, 2012b) as dependent from temperature. The graphite elements and the fuel elements are characterized by the properties reported in Table 2, the materials are considered to be isotropic. The control rods are modelled as fuel elements without a power source. Such choice is justified by the fact that the temperature of the control rods is not of interest for this work.

| Constant symbol | Value | Unit |
|---|---|---|
| $c_p$, fuel elements | 365 | kJ kg$^{-1}$K$^{-1}$ |
| $\lambda$, fuel elements | 16.3 | W m$^{-1}$K$^{-1}$ |
| $\rho$, fuel elements | 6300 | kg m$^{-3}$ |
| $c_p$, graphite | 710 | J kg$^{-1}$K$^{-1}$ |
| $\lambda$, graphite | 110 | W m$^{-1}$K$^{-1}$ |
| $\rho$, graphite | 1950 | kg m$^{-3}$ |

*Table 2 - Thermal properties used in the model, for solids.*

## 4. SUB-COOLED BOILING MODELLING

It was experimentally proved by Mesquita et al. (2006) that sub-cooled boiling conditions can be reached on some fuel elements. Adopting the model as above formulated, a large part of the inner fuel elements surfaces reach temperatures far above the water saturation temperature, when total power exceeds 100 kW. As a consequence, the fuel temperature is not properly simulated. The model has therefore been extended to take into account such phenomenon.

Several CFD approaches exist that allow modelling boiling heat transfer. However, such approaches are generally based on a two-fluid description, so that their adoption would result in a great increase of the model complexity and its computational requirements.

On the other hand, forced convection boiling can be addressed by *heat partition modelling*, separating the different contributes to the heat flux. An example of this approach can be found in the largely adopted Chen correlation (Chen, 1966), that superimposes convective and boiling contributions to determine the heat transfer coefficient:

$$h_{tot} = F\, h_l + S\, h_{pb} \tag{13}$$

Here, $h_{tot}$ is the heat transfer coefficient, S is a suppression factor, F is a factor representing the increase in convective heat extraction due to two phase flow. The two heat transfer coefficients $h_l$ (liquid) and $h_{pb}$ (pool boiling) are determined from other correlations, specific to the heat transfer mode. In Chen, the internal flow into tubes is addressed by means of the Dittus-Boelter (Dittus and Boelter, 1930) correlation for single phase flow and the Forster-Zuber (Forster and Zuber, 1955) for the pool boiling part. The two correcting factor are, instead, determined as a function$^{1}$ of the vapour quality and the flow regime. The Chen correlation can be effectively extended to sub-cooled boiling range by setting F equal to 1 and leaving the S factor to be dependent from the Reynolds number only (Rohsenow et al., 1998). In short, the Chen correlation is not dependent from the vapour conditions when addressing sub-cooled boiling. Therefore, the choice of not simulating the vapour as a second fluid is further justified.

The adoption of the Chen correlation or an analogous one cannot be effectively used in CFD because the definition of the bulk flow Reynolds number becomes extremely arbitrary for complex geometries, resulting, for example, in a poor definition of the suppression factor.
In our model, we decided to use a pool-boiling correlation to determine the boiling heat flux, neglecting the suppression effects. Such choice is justified by the fact that S would be roughly equal to 0.95 with the typical Reynolds number in the TRIGA core. The absence of something similar to the Chen suppression factor is likely to result in a slight underestimation of the wall temperature for the analysed operating point.
From the point of view of the adopted model, the boiling effects are treated as a boundary condition on the fuel-fluid interface. In particular, the heat flux at the boundary is determined as:

$$q_w'' = -\lambda \nabla T_w = q_{source}'' - q_{pb}'' \qquad (14)$$

where the source contribution is due to the generated power into the fuel element, while the boiling contribution to the heat flux is determined with the Rohsenow correlation (Rohsenow, 1952):

$$q_{pb}'' = h_{fg}\,\mu_l\left(\frac{g(\rho_l - \rho_v)}{\xi}\right)^{1/2}\left(\frac{c_{p,l}(T_w - T_{sat})}{C_{sf}\,H_{fg}\,Pr^{n_{fluid}}}\right)^3 \qquad \text{if } T_w > T_{sat} \qquad (15)$$

where $C_{sf}$ is a constant relative to the coupling of surface material and cooling fluid. $C_{sf}$ is equal to 0.011 for rounded aluminium plates and water (Pioro, 1999). The exponent $n_{fluid}$ is equal to 1 for water. The vapour and liquid thermal properties are evaluated at saturation temperature. Equation (14) is substituted in equation (B.7) in order to determine the wall temperature. The choice of the Rohsenow correlation instead of the Forster-Zuber one, generally adopted in the Chen model, is due to the fact that the former takes into account the material coupling, while the latter does not.
The adopted modelling does not give information on the energy transfer from the vapour to the water and the bubble quenching on the surface. However, considering the low boiling regime, the induced error is considered to be far lower than the RANS modelling uncertainties. As far as neutronics is concerned, we performed a study to evaluate how much the uncertainty on the water temperature can influence reactor criticality (see section 6.1).

## 5. TERMAL-HYDRAULIC RESULTS

The model equations have been discretized and solved with the Finite Elements approach, using linear Lagrange elements in order to maintain the degree of freedoms into an acceptable range. The mesh, consisting of about $2 \cdot 10^6$ elements, is depicted in Figure 4. The GMRES solver, with a SSOR vector left preconditioner has been selected for the solution of the stationary problem (Comsol, 2012b).

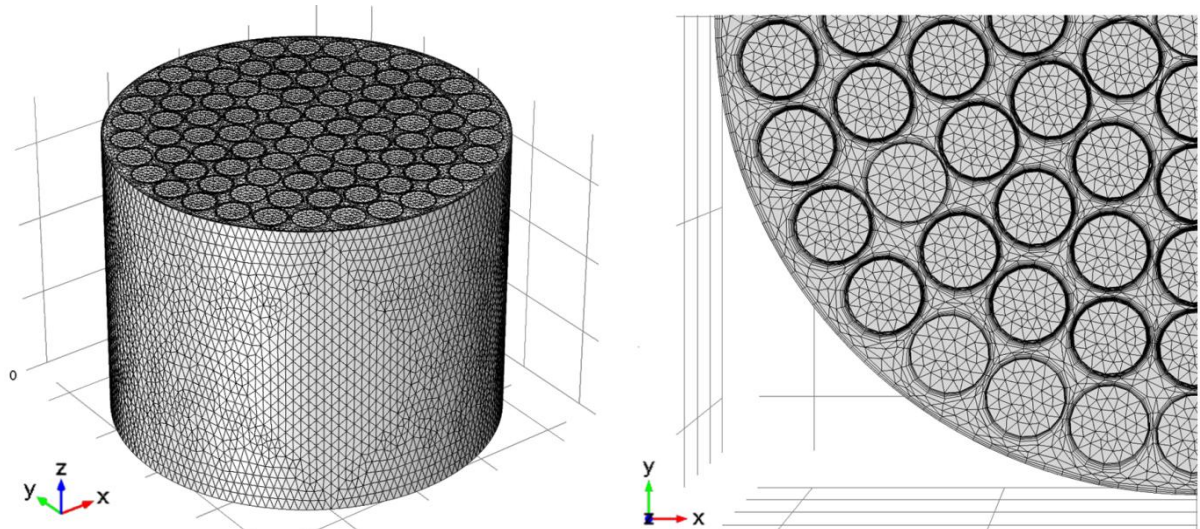

**Figure 4 - Complete mesh and particular from the top surface**

The results regarding the solution of the velocity field are shown in figures 5, 6, 7. Figure 5 shows the velocity magnitude distribution in all the fluid volume, only half of the geometry is represented for visualization reasons. In figure 6 a 3-D visualization of the flow is reported at different heigths. If figure 7 the flow velocity on the x-y plane is depicted for different heights.

Considering both figures 5 and 6, the effects of bouyancy on the velocity field are evident in the upper part of the core. Since in the core center the temperatures are higher than in the outside regions, the fluid velocity increases in the inner region. Moreover, in the outer ring, the water velocity is higher near the fuel elements than near the unheated graphite ones. Due to the imposition of an uniform velocity at the inlet, the cross-flow between fuel elements is dominated by the axial component of the flow. However, as shown in figure 7, a slow inward motion of the fluid is detectable. This is consistent with the increase of the velocity in the core center. In the lower part of the core, such effects are negligible. In the upper part, instead, the "x-y plane" water velocity reaches the 6% of the total velocity magnitude, on average, with peaks up to 13%. As it can be observed, the cross-flow is directed to the zone surrounding the B ring, resulting in a larger flow rate in that region, as observed in figure 6.

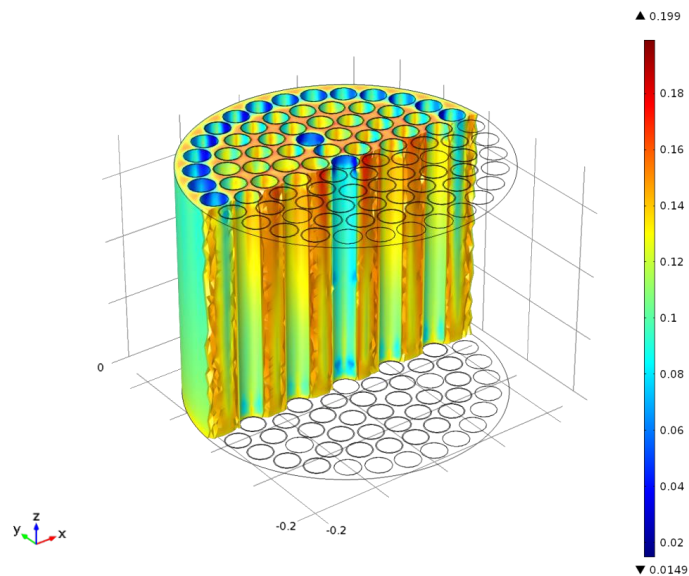

**Figure 5 - Flow magnitude distribution in m/s, half core.**

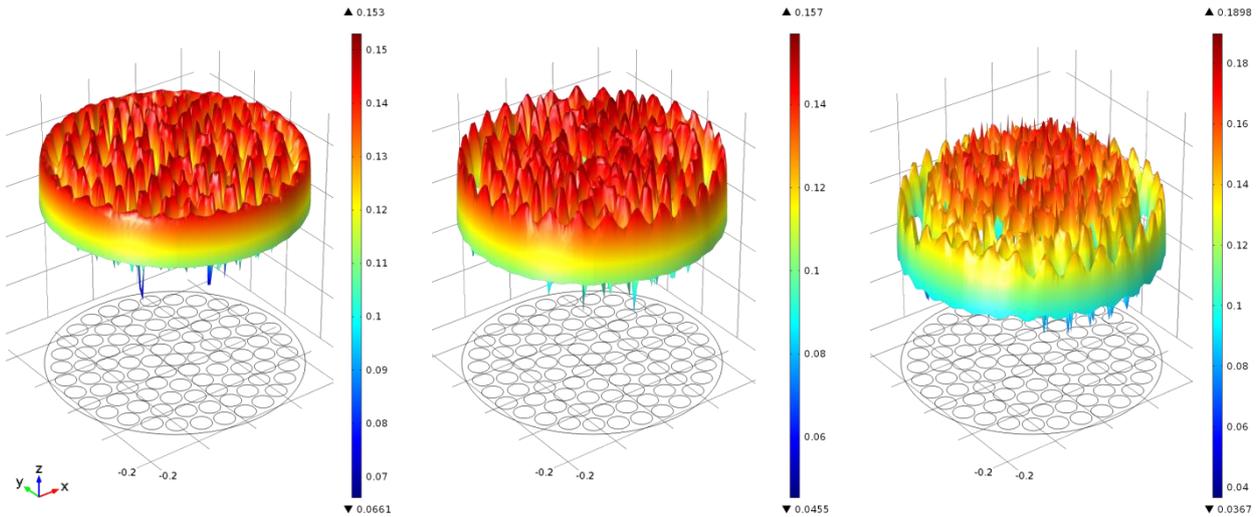

**Figure 6 - Flow velocity magnitude distribution in m/s, on x-y planes from left to right: sections taken at 6, 18 and 30 cm above the geometry bottom. Velocity at the wall (equal to zero) is not shown. Lift-off velocity (Appendix B) is shown instead.**

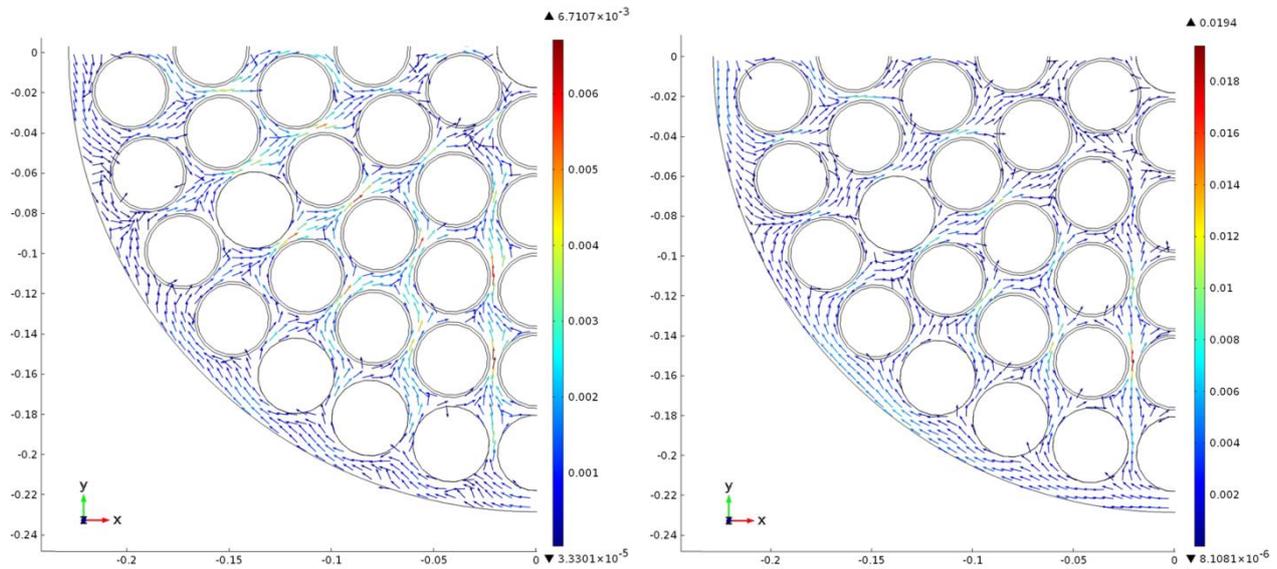

**Figure 7 - Longitudinal flow distribution on x-y planes. From left to right: sections taken at 6 and 30 cm above the bottom. The colour reflects the value of the magnitude of the longitudinal components in m/s.**

Energy equation solutions are reported in figures 8, 9 and 10. In particular, the fuel temperature in 3-D geometry is reported in figure 8; the temperature profile on a vertical section for different heights is reported in figure 9; the fuel element wall temperature, as well as the regions in which the model predicts boiling are shown in figure 10. The temperature distribution is characterized by lower values in the most peripheral parts of the core, while higher ones are reached in the central region, in which the power density is larger. The temperature maximum is located above the mid-plane of the most inner fuel elements. In the lower region of the core, the clad temperature undergoes a steep increase from the core bottom to the saturation region, ranging from the inlet temperature to the saturation one (111.3 °C at 1.5 bar). In the region in which saturation temperature is reached, sub-cooled boiling starts to contribute to heat transfer, so that the increase of the clad temperature above the saturation point is only in the order of few degrees, thanks to better heat removal. In the top part of the core, the heat flux is not high enough to maintain the temperature above saturation point.

The different spacing between the fuel elements produces some non-uniformity on the surface temperature of each element. This effect is clearly recognizable from figure 10: in the zone in which elements of the same ring are closer, the temperature is lower than in the other parts of the same element, given the same height, because fluid-dynamics effects lead to better heat removal (see figure 7). As an example, consider elements in the D, E, F rings. The surface temperature on E element walls is above the saturation point where opposed to D and E ring elements, but the temperature on the zones where the E elements are opposing each other is lower.

The presence of sub-cooled boiling heat transfer allows a certain "stabilization" of the average fuel temperature with respect to fluid temperature variations at the core inlet. The relation between the heat flux and the temperature is quasi-linear in forced convection heat transfer. If only this phenomenon is considered, an increase in the inlet temperature will result in a similar increase of the fuel surface temperature, and therefore in an analogous increase of the fuel temperature, being that the heat exchange is only slightly increased by the reduction of the viscosity. On the other hand, pool boiling correlates the heat flux to the third power of the difference between the wall temperature and the saturation point (see equation (15)). Consequently, an increase in the inlet temperature would result in a "faster" reaching of the saturation temperature, but the temperature excess would then be mitigated by the sub-cooled boiling.

If an increase of 10 °C at the inlet temperature is considered, the wall temperature rise in the boiling zone is about 2°C. In the non-boiling zone the increase is linear. Moreover, the boiling zone expands, so that the fuel temperature increase is there less than linear. Thus, the increase in the average fuel temperature of the whole reactor is about 5.6 °C. Those effects are shown in figure 11. In short, boiling allows the effects of a variation of the inlet temperature to be dumped from the point of view of the fuel temperature.

The sensitivity of the fuel temperature with respect to the variation of the mass flow rate has been investigated by reducing the inlet flow by 10% (i.e., imposing a flow-rate of about 8.4 kg/s). As shown in Figure 12 the effect is strongly mitigated in the boiling region, while the temperature increases elsewhere due to less efficient heat removal. The average fuel temperature is increased by 0.7 °C.

Experimental temperature values are available from the instrumented fuel elements positioned in the B ring (Cambieri et al., 1965). The instrumented elements are provided with three thermocouples positioned at the half of the active height and one inch under and above that position. Considering the radial temperature increase in those fuel elements, that varies from 200 to 287 °C in a 1.8 cm wide region, the exact position of the thermocouple should be known for a consistent comparison of simulated and experimental data. According to (Tomsio, 1986) the sensing tips of the thermocouples are located about two-thirds of the distance between the outer radius and the vertical centerline of the elements, but no more information are provided. Considering the modelling hypothesis and the scattered experimental data, showing a mean of 253.3 °C and a standard deviation of 9.9 °C at the fuel midplane, the simulation of the temperature field can be considered successful. For comparison, in figure 13 the midplane section of a fuel element of the B ring is shown. This result gives a good indication that the adopted modelling approach is able to reproduce realistic core conditions. In particular, sub-cooled boiling regime is confirmed. Indeed, considering the results presented in figure 14, the fuel temperature would be more than 30 K higher in the considered section if no boiling was established. Such condition does not agree with experimental data.

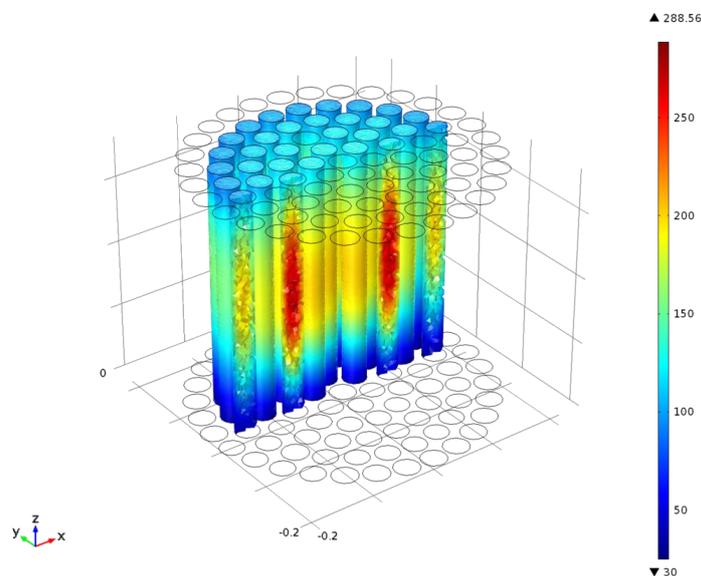

**Figure 8 - Fuel temperature distribution in °C, half core. The clad outer temperature is not shown here.**

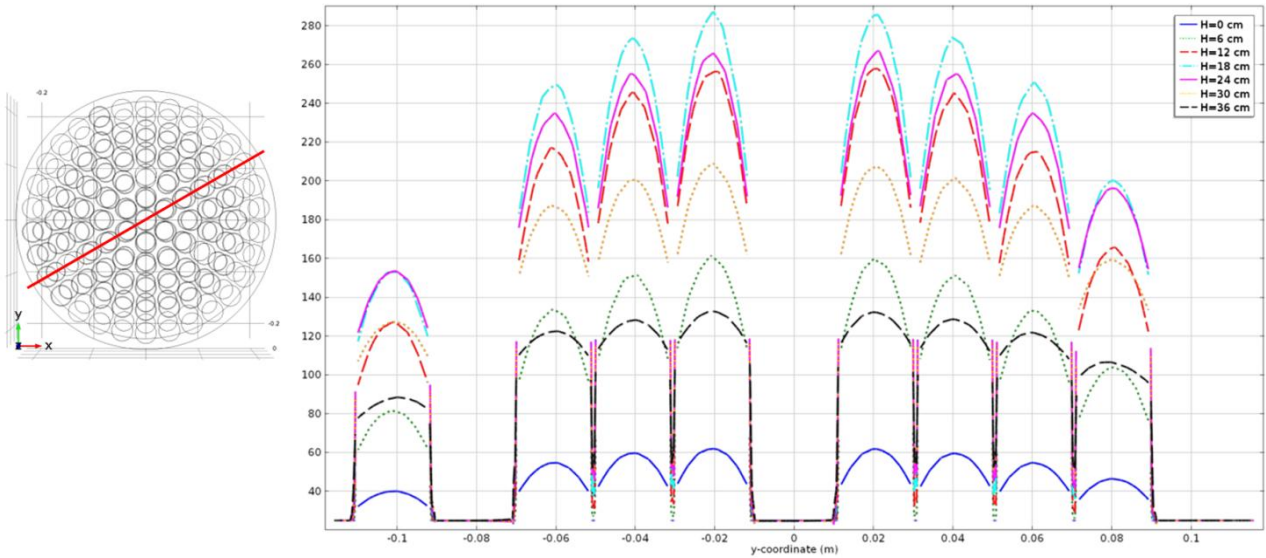

**Figure 9 - Fuel temperature, in °C. The discontinuity in the temperature profile corresponds to the cladding-gap regions.**

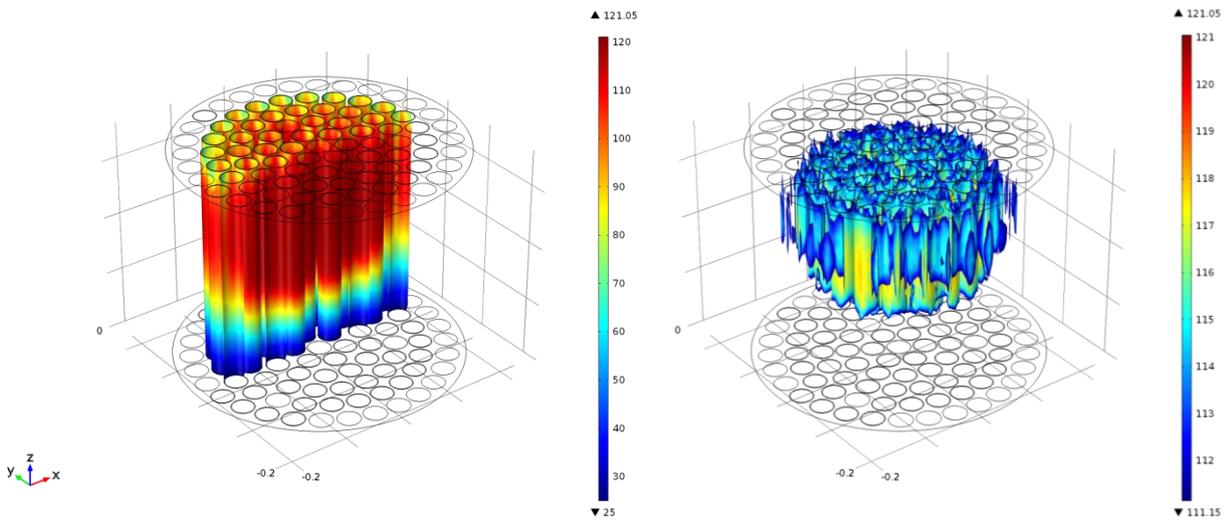

**Figure 10 - Fuel element wall temperature, half core in °C. In the right picture, the region above the saturation temperature at 1.5 bar (111.3 °C) is shown.**

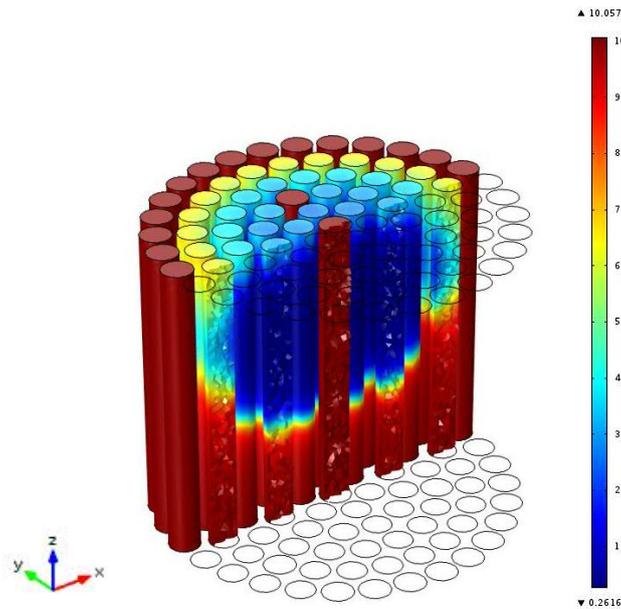

**Figure 11 - Fuel temperature distribution in °C, half core: increment due to the increase of the inlet temperature from 25 to 35 °C. The clad outer temperature is not shown here.**

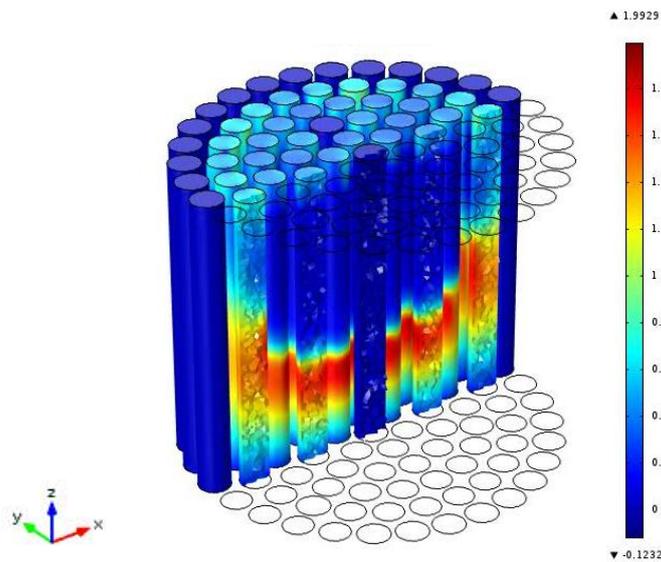

**Figure 12 - Fuel temperature distribution in °C, half core: increment due to the decrease of mass flow by 10%. The clad outer temperature is not shown here.**

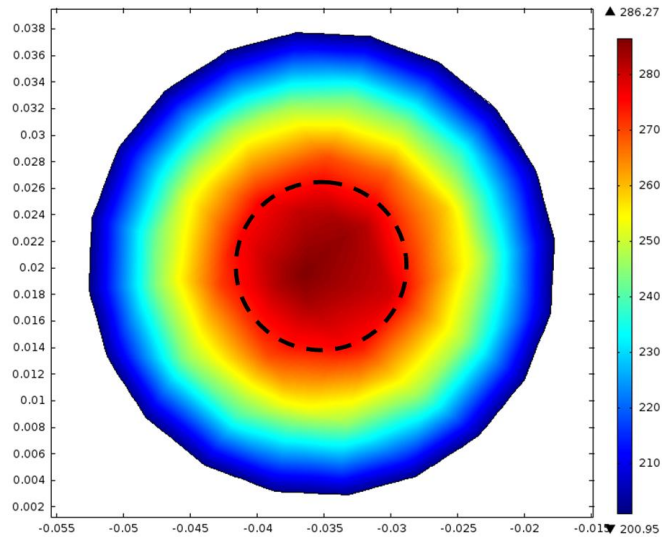

**Figure 13 - Temperature distribution in °C, on the midplane of a fuel element placed in the B ring. The dashed line indicates the supposed position of thermocouple tip.**

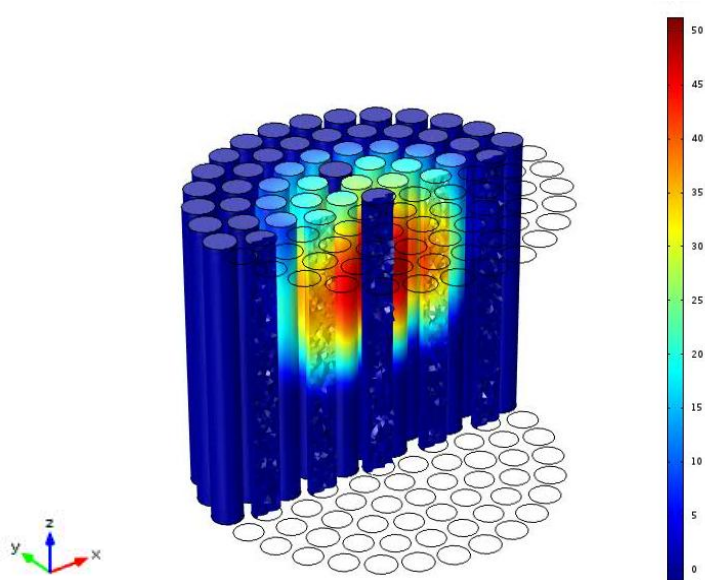

**Figure 14 - Fuel temperature distribution in °C, half core: increment due to the neglecting the boiling effects. The clad outer temperature is not shown here.**

## 6 THE MCNP MODEL FOR NEUTRONICS SIMULATIONS AT FULL-POWER

The temperature distribution obtained from the thermal-hydraulic calculations, was finally introduced in the MCNP reactor model to simulate the full-power steady state. A discretization of the thermal map was needed to assign cross sections at different temperatures to the volumes describing the fuel. The fuel elements were subdivided into 5 sections (each 7.2 cm high) and the average temperature was determined for each section, depending on the occupied position along the vertical and radial core axes.

Since the ENDF/B-VII cross sections are provided for temperatures at each 50 or 100 °C, which are steps too coarse for a good description of the thermal distribution, we used the MAKXSF code (Brown, 2006) to generate new cross sections, interpolated at each 10 °C, for ZrH, uranium and $H_2O$.

In this way, it has been possible to approximate the average temperature of each fuel section to the closest one available in the new cross section list (Tab. 3), allowing a more accurate description of the thermal effects which influence the low-energy scattering on ZrH and the interactions with uranium.

| Fuel section | Core Ring | | | | |
|---|---|---|---|---|---|
| | B | C | D | E | F |
| 1 | 430 | 420 | 410 | 390 | 380 |
| 2 | 490 | 480 | 460 | 430 | 400 |
| 3 | 500 | 500 | 480 | 430 | 400 |
| 4 | 480 | 460 | 440 | 400 | 370 |
| 5 | 370 | 360 | 350 | 360 | 330 |

*Table 3: Temperature values (in K) employed to model the thermal distribution in the full-power MCNP simulations. The fuel sections are numbered 1 → 5 from top to bottom.*

The resulting full-power MCNP model was then benchmarked using the data available in the First Criticality Final Report (Camberi et al., 1965) concerning four criticality configurations at 250 kW power, with the control rods in different positions.
Experimentally, the decrease of reactivity due to the thermal effects is compensated by withdrawing the control rods: using the calibration curves and comparing the criticality positions at full-power with those at zero-power, the experimental reactivity loss has been estimated to be (1.36 ± 0.06)$.

Similarly, in the MCNP simulations, we can check that the variation induced on $k_{eff}$ by changing the temperature from 294 K to the full-power distribution is properly counterbalanced by simulating the control rods in the positions experimentally recorded at 250 kW. In practice, we took as reference point the average value of reactivity obtained from the simulations of the zero-power criticality configurations (Alloni et al., 2014) and we verified that the reactivity values from the full-power simulations resulted compatible again with the 0 $ criticality condition.
The simulation results are presented in Figure 15 and Table 4, where a $\beta_{eff}$ parameter equal to 0.0073 was used to express the simulation results in $ units (Cambieri et al., 1965). The shadowed area corresponds to the standard deviation of the results obtained from the simulations of different zero-power criticality configurations ($1\sigma = 0.077$ $). This uncertainty bandwidth includes the systematic errors - occurring when changing the control rod positions - which affect the MCNP simulation model, due to some lack of information about the materials and the geometric details of some core components.

Taking into account this systematic uncertainty, the results obtained with the full-power reactor model are in good agreement with the 0 $ criticality condition: indeed, the average value of the results is −0.082 $ with a standard deviation of 0.084 $. Therefore, we can state that the thermal effects, which cause a reactivity decrease of about 1.36 $, are properly simulated in the full-power MCNP model with a precision of about 6%.

Considering the average of the fuel temperature corresponding to the situation described in table 3, equal to 147.65 °C, the fuel temperature feedback coefficient can be determined equal to -(0.01072 ± 0.00047) $/°C, using zero-power as reference condition. Such value is extremely similar to the one obtained in (Cammi et al., 2013) by fitting transient data. This is a further verification of the methodology adopted in this work.

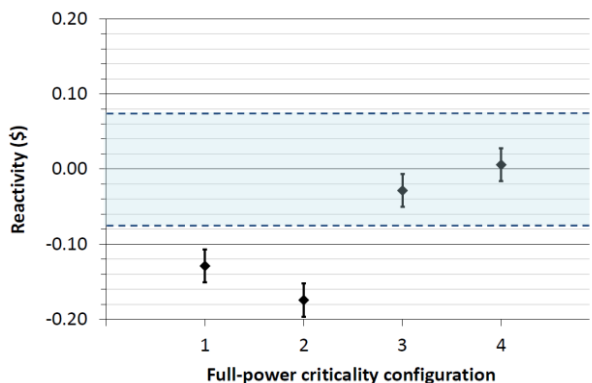

| Cfg. number | CR Position (digit) | | Reactivity ($) |
|---|---|---|---|
| | SHIM | REG | [±0.02$_{stat}$] |
| 1 | 524 | 818 | -0.13 |
| 2 | 556 | 605 | -0.17 |
| 3 | 588 | 557 | 0.03 |
| 4 | 673 | 350 | 0.01 |

**Figure 15.and Table 4 - Reactivity values of the full-power reactor simulations with the control rods in the experimental criticality configurations measured in 1965 (the TRANSIENT rod was completely withdrawn).**

In order to separately quantify the thermal effects due to the heating of ZrH and uranium, some simulations were run using the cross sections of $^{235}$U and $^{238}$U at 294 K, while maintaining the cross sections of ZrH corresponding to those listed in Table 3.

The results highlight that the main contribution to the reactivity loss is due to ZrH: in fact, the thermal effect due to uranium heating has been estimated equal to $(-0.20 \pm 0.02)$ \$, in front of an overall thermal effect of about $-1.36$ \$.

The thermal effects concerning the heating of uranium, which are related to the Doppler broadening of the absorption resonances, were then investigated in more detail by separately simulating $^{235}$U or $^{238}$U at room temperature. This analysis showed that the heating of $^{235}$U induce a positive reactivity change equal to $(0.17 \pm 0.02)$ \$, while the heating of $^{238}$U causes a reactivity loss equal to $(-0.41 \pm 0.02)$ \$. It is worth noting that the sum of these two effects is compatible with the overall reactivity change due to uranium previously estimated.

Since the thermal effects are mainly caused by a lower effectiveness of ZrH in moderating the neutrons, the flux spectra in the fuel volumes obtained from zero-power and full-power simulations were compared. Looking at the spectra in Figure 15, which refer to the fuel volumes in the central region of the core, it is interesting to note that, at full-power condition, the thermal spectrum component is shifted towards higher energy values. Moreover, to compensate for the reactivity loss, the intensity of the neutron flux (normalized per simulated neutron) is higher when the fuel is simulated in the hot condition.

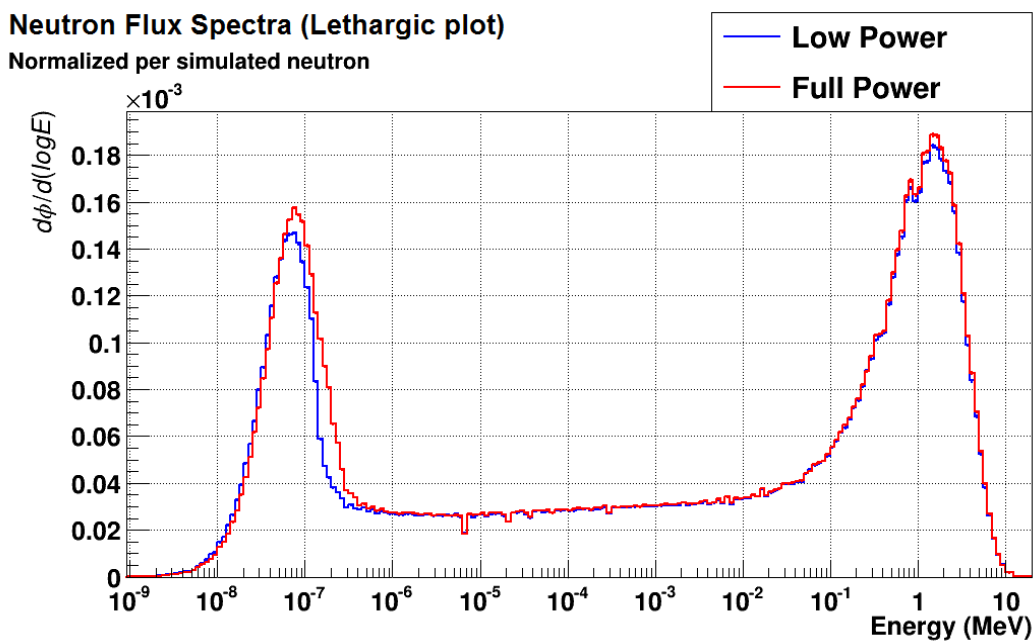

**Figure 16** -Comparison of the neutron flux spectra obtained from zero-power and full-power simulations in the fuel volumes at the central region of the core (the error bars are very little because the MCNP simulation was run with high statistics).

**6.1 Analysis of the water thermal effect**

Finally, since the water in the core is heated when the reactor operates at full-power steady state, the thermal effects related to water were also investigated. In this case, two effects must be considered:
- the different moderating effectiveness of water, which can be simulated by using the $S(\alpha,\beta)$ cross sections of hydrogen in $H_2O$, available for different temperatures;
- the variation of water density.

As stated in section 5, the thermal-hydraulic model results on water temperature are strongly dependent from boundary conditions. Therefore, the analysis has been performed on a wide range of water temperatures. This also allows to evaluate the water effect in a more general way. In order to analyze the dependence of reactivity versus water temperature, different simulations were run, covering the temperature range between 21°C and 77°C (while the fuel temperatures were maintained at the values listed in Table 3). Particularly, we performed three sets of simulations: in the first, both the cross section and the density were varied; in the second, only the cross section was changed; in the third, only the water density was modified (Figure 16). In this way, it was possible to analyze the thermal effect of each component, obtaining the following results:
- positive reactivity variation related to the use of $S(\alpha,\beta)$ cross sections at higher temperatures;
- negative reactivity variation due to the density decrease.

The overall effect results in a slight positive variation of reactivity.

Due to the presence of the vapour bubble in the reactor core, the average density of the water is slightly lower than in the corresponding fluid-only case, further reducing the reactivity increment. Since the average temperature in the core is not expected to exceed 50°C, we can state that the water heating causes a reactivity variation of +0.2 $ at maximum. If this effect is included in the full-power model, the results would be centred at 0.12 $, being again compatible with the expected 0 $ criticality condition (taking into account the systematic uncertainty of 0.08 $ affecting the simulation model). If we consider that the real water temperature in the reactor core is within the two simulated temperature extreme conditions, the agreement between the simulation and the experimental data is very good.

For a better simulation of the thermal effects related to water, a more refined thermal-hydraulic model will be developed in the future for evaluating the temperature distribution of water in the core. Anyway, at the moment, considering the uncertainties affecting the simulation model, we can affirm that the reactivity variation due to thermal effects is correctly simulated in MCNP model developed for the full-power TRIGA reactor.

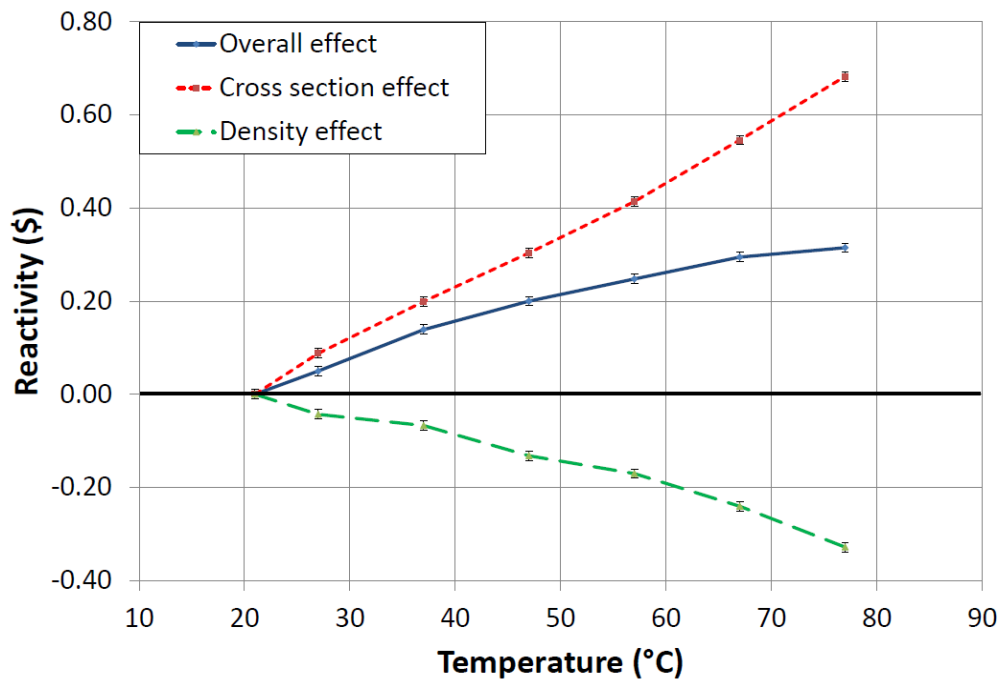

**Figure 16 - Reactivity variation as function of water temperature: the continuous (blue) line interpolates the results from simulations where the density and the cross section were both changed; the dotted (red) line refers to simulations where only the water cross section was varied; the dashed (green) line refers to simulations where only the water density was modified.**

**CONCLUSIONS**

In this work, the modelling of temperature reactivity effects in the TRIGA Mark II of the University of Pavia is performed by means of the coupling of the results obtained with a MC code for neutronics and a "Multiphysics code" for thermal-hydraulics.
The thermal-hydraulic model reproduces the active core region, solving both fluid and energy equations. The modelling of the core alone requires the use of an imposed flow rate approach, instead of a more realistic free circulation one that would require the modelling of the whole reactor pool. Nevertheless, some of the buoyancy effects are taken into account. Due to the turbulent nature of the flow in the core, a RANS $k$-$\omega$ model is adopted for the simulation. Moreover, sub-cooled boiling effect is taken into account by means of a boundary condition based on the Rohsenow correlation. It is shown that when the reactor is at full-power the average fuel temperature is only slightly influenced by small variations of the core inlet temperature and flow rate, thanks to the sub-cooled boiling regime. The obtained results are coherent with the available experimental data.
The MC simulations performed by adopting the full-power temperature distribution, as determined by the thermal-hydraulic model, show good agreement with the experimental data. Indeed, the reactivity defect caused by the temperature increase with respect to room temperature is correctly reproduced. Finally, it is shown that an increase in the water temperature leads to a small but positive reactivity contribution.

# APPENDIX A

This appendix deals with definitions of terms and closure equations for the model presented by equations (3-5) The turbulent kinetic energy production $P_k$ term is defined as:

$$P_k = \mu_T \nabla \mathbf{u} : \left((\nabla \mathbf{u}) + (\nabla \mathbf{u})^T\right) - \frac{2}{3}\left(\mu_T (\nabla \cdot \mathbf{u})^2 + \rho k \nabla \cdot \mathbf{u}\right) \tag{A.1}$$

The turbulent Prandtl number is determined by means of the Kays-Crawford model (Comsol, 2012a), so that the turbulent thermal conductivity can be determined:

$$\Pr_T = \left(\frac{1}{2\Pr_{T\infty}} + \frac{0.3}{\sqrt{\Pr_{T\infty}}}\frac{c_p \mu_T}{\lambda} - \left(0.3\frac{c_p \mu_T}{\lambda}\right)^2 \left(1 - e^{-\frac{\lambda}{0.3 c_p \mu_T \sqrt{\Pr_{T\infty}}}}\right)\right)^{-1} \tag{A.2}$$

the parameter $\Pr_{T\infty}$ is set to 0.85.

Finally, the closure coefficients are abridged in Table A.1, the β parameters are determined from the $(-)_0$ ones depending on the velocity field, in particular, they are multiplied for a function of the average rotation-rate tensor and the mean strain-rate tensor.

| Symbol | Value |
| --- | --- |
| $A$ | 13/25 |
| $\sigma_k^*$ | 1/2 |
| $\sigma_\omega$ | 1/2 |
| $\beta_0^*$ | 9/125 |
| $\beta_0$ | 9/100 |

*Table A.1- Closure coefficients for k-ω equations.*

# APPENDIX B

The wall functions used in COMSOL are such that the computational domain is assumed to start at a distance $\delta_w$ (wall lift-off) from the wall, as shown in Figure B.1. Such approach is generally known as "scalable wall function" and allows using not extremely refined meshes near the walls, so that it is suited to "industrial" codes but not for investigating the flow in the viscous sub-layer. The velocity field is modelled with a logarithmic law in the viscous sub-layer.

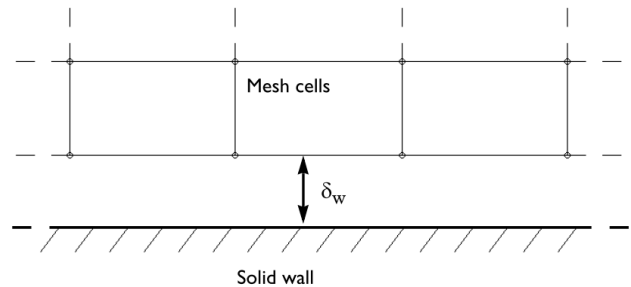

**Figure B.1 - Wall treatment adopted in COMSOL**

The boundary conditions at the wall are stated as follows, with a no-penetration condition (B.1) and a shear stress condition (B.2) for the velocity field, a homogeneous Neumann condition for the turbulent kinetic energy (B.3) and a Dirichlet condition for the specific dissipation rate (B.4):

$$\mathbf{u} \, \hat{\mathbf{n}} = 0 \tag{B.1}$$

$$\left(-p\mathbf{I} + (\mu + \mu_T)((\nabla \mathbf{u}) + (\nabla \mathbf{u})^T) - \frac{2}{3}\rho k\,\mathbf{I}\right)\hat{\mathbf{n}} = -\rho\frac{u_\tau}{\delta_w^+}(\mathbf{u} - (\mathbf{u}\cdot\hat{\mathbf{n}})\hat{\mathbf{n}}) \tag{B.2}$$

$$\nabla k\,\hat{\mathbf{n}} = 0 \tag{B.3}$$

$$\omega = \rho\frac{k}{\kappa_v\delta_w^+\mu} \tag{B.4}$$

where the wall lift-off in viscous units $\delta_w^+$ and the wall velocity are defined as:

$$\delta_w^+ = \rho\frac{k^{1/2}\delta_w}{\mu}\beta_0^{*\,1/4} \tag{B.5}$$

$$u_\tau = \frac{|\mathbf{u}|}{\frac{1}{\kappa_v}\ln(\delta_w^+) + B} \tag{B.6}$$

in which B is a constant, set to 5.2, and $\kappa_v$ is the Von Karman constant, equal to 0.41.
The software determines the wall lift-off $\delta_w$ so that $\delta_w^+$ is equal to 11.06. The $\delta_w$ is lower bounded, therefore values of $\delta_w^+$ above the objective one can be achieved. The accuracy of the solution has to be verified via two different criteria: i) the $\delta_w^+$ value required by the software on the wall is fulfilled; ii) the computed value of $\delta_w$ is small with respect to the surrounding dimensions of the geometry. Whether one of this conditions is not satisfied in some part of the geometric domain, the mesh has to be refined. The adopted wall treatment approach requires a thermal wall function to be used, in order to determine the correct wall temperature $T_w$, as opposed to the fluid temperature $T_f$ computed on the meshed wall boundary. The heat flux on the fictitious $\delta_w$ width is defined as:

$$q_{fw}'' = \rho\frac{k^{1/2}\beta_0^{*\,1/4}c_p}{T^+}(T_w - T_f) \tag{B.7}$$

where $q_{fw}''$ is set equal to the thermal flux prescribed on the boundary. The dimensionless temperature $T^+$ is defined as:

$$T^+ = \begin{cases} \Pr\delta_w^+ & \text{for } \delta_w^+ < \delta_{w1}^+ \\[6pt] (15\Pr^{2/3} - 500/\delta_{w2}^+) & \text{for } \delta_{w1}^+ \leq \delta_w^+ < \delta_{w2}^+ \\[6pt] \frac{\Pr}{\kappa_v}\ln(\delta_w^+) + 15\Pr^{2/3} - \frac{\Pr_T}{2\kappa_v}\left(1 + \ln\left(1000\frac{\kappa_v}{\Pr_T}\right)\right) & \text{for } \delta_{w2}^+ \leq \delta_w^+ \end{cases} \tag{B.8}$$

where the two decision variables are:

$$\begin{cases} \delta_{w1}^+ = 10/\Pr^{1/3} \\[6pt] \delta_{w2}^+ = 10\sqrt{10\,\kappa_v/\Pr_T} \end{cases} \tag{B.9}$$

**Nomenclature**

*Latin symbols*

| | |
|---|---|
| A | closure coefficient for k-ω modelling, - |
| $c_p$ | heat capacity at constant pressure, J kg$^{-1}$ K$^{-1}$ |
| $C_{sf}$ | surface-fluid coupling constant, - |
| $d_s$ | gap-cladding width, m |
| E | energy, eV |
| F | convective heat extraction factor, - |
| **g** | gravity field, m s$^{-2}$ |
| h | heat transfer coefficient, W m$^{-2}$ K$^{-1}$ |
| $H_{fg}$ | enthalpy of vaporization, J kg$^{-1}$ |
| **I** | identity tensor, - |
| k | turbulent kinetic energy, m$^2$ s$^{-2}$ |
| $k_{eff}$ | multiplication factor, - |
| $L_T$ | turbulence length scale, m |
| N | number of atoms, - |
| $\hat{\mathbf{n}}$ | versor, - |
| $n_{fluid}$ | fluid constant in Rohsenow correlation, - |
| p | pressure, Pa |
| $P_k$ | turbulent kinetic energy production, W m$^{-3}$ |
| Pr | Prandtl number, - |
| $Pr_T$ | turbulent Prandtl number, - |
| $Pr_{T\infty}$ | turbulent Prandtl number parameter, - |
| q" | heat flux, W m$^{-2}$ |
| $R_f$ | average fission rate, m$^{-3}$ s$^{-1}$ |
| S | suppression factor, - |
| T | temperature, K |
| t | time, s |
| $T^+$ | dimensionless temperature (Wall function), - |
| **u** | velocity field, m s$^{-1}$ |
| $U_0$ | inlet velocity, m s$^{-1}$ |

*Greek Symbols*

| | |
|---|---|
| $\beta_{eff}$ | effective delayed neutron precursor fraction, - |
| $\beta_0$ | closure coefficient for k-ω modelling, - |
| $\beta_0^*$ | closure coefficient for k-ω modelling, - |
| $\kappa_v$ | Von Karman constant, - |
| $\delta_w$ | wall lift-off, m |
| $\delta_w^+$ | wall lift-off in viscous units, - |
| $\boldsymbol{\lambda}$ | thermal conductivity, W m$^{-1}$ |
| $\lambda_s$ | gap-cladding thermal conductivity, W m$^{-1}$ |
| $\mu$ | dynamic viscosity, Pa s |
| $\mu_T$ | turbulent dynamic viscosity, Pa s |
| $\xi$ | surface tension, N m$^{-1}$ |
| $\rho$ | density, kg m$^{-3}$ |
| $\sigma_k^*$ | closure coefficient for k-ω modelling, - |
| $\sigma_\omega$ | closure coefficient for k-ω modelling, - |
| $\sigma$ | standard deviation, - |
| $\Sigma_f$ | microscopic fission cross section, b |
| $\varphi$ | neutron flux, m$^{-2}$ s$^{-1}$ |
| $\omega$ | specific dissipation rate, s$^{-1}$ |

**Subscripts**

| | |
|---|---|
| *f* | fluid |
| *fw* | fluid-wall |
| *l* | liquid |
| *pb* | pool boiling |
| *sat* | saturation |
| *tot* | total |
| *v* | vapour |
| *w* | wall |

**Acronyms**

| | |
|---|---|
| 3-D | Three-Dimensional |
| GMRES | Generalized Minimal Residual Method |
| MC | Monte Carlo |
| MCNP | Monte Carlo N-Particle |
| RANS | Reynolds Averaged Navier-Stokes |
| REG | regulating rod |
| SHIM | shim rod |
| SSOR | Symmetric Successive Over Relaxation |
| TRANS | transient rod |
| TRIGA | Training Research and Isotope production General Atomics |